\documentclass[journal]{IEEEtran}

\usepackage[dvips]{graphicx}

\graphicspath{{figures/}}
\DeclareGraphicsExtensions{.eps}

\usepackage[cmex10]{amsmath}

\usepackage{amssymb}

\usepackage{cite}

\begin{document}

\title{Probing the sheath electric field with a crystal lattice by using thermophoresis in dusty plasma}

\author{Victor~Land, Bernard~Smith, Lorin~Matthews, Truell~Hyde~\IEEEmembership{Fellow,~IEEE}
\thanks{V. Land, B. Smith, L. Matthews, and T. Hyde are with the Center for Astrophysics, Space Physics and Engineering
Research, at Baylor University, Waco, TX, 76798-7316 USA, e-mail: victor\_land@baylor.edu,  (see http://www.baylor.edu/CASPER).}
\thanks{Manuscript received xxxxx xx, 2009; revised xxxxx xx, 2009.}}

\markboth{IEEE transactions on plasma science,~Vol.~XX, No.~XX, XXXXX~2010}%
{Land \MakeLowercase{\textit{et al.}}: Dust particles as probes for the sheath electric field by applying additional
thermophoresis}

\maketitle

\begin{abstract}
A two-dimensional dust crystal levitated in the sheath of a modified Gaseous Electronics Conference (GEC) reference cell is manipulated
by heating or cooling the lower electrode. The dust charge is obtained by measuring global characteristics of the levitated
crystal obtained from top-view pictures. From the force balance, the electric field in the sheath is reconstructed. From the
Bohm criterion, we conclude that the dust crystal is levitated mainly above and just below the classical Bohm point.
\end{abstract}


\IEEEpeerreviewmaketitle

\section{Introduction}

\IEEEPARstart{D}{ust} particles present in laboratory plasma collect electrons and ions and become negatively charged. The
charge on a particle is estimated to be between 1 and 5 electron charges per nanometer radius \cite{Stoffels1999}, depending on the background
pressure \cite{Konopka1997,Ratynskaia2004}. Therefore, micrometer-sized particles used in dusty plasma experiments can carry a negative charge of
thousands of electron charges. Because of this, the dust particles are levitated against gravity in the strong electric
fields present in the sheath. The electrostatic interaction between the particles is shielded by the plasma, but depending on
the discharge parameters, strongly coupled systems can be formed \cite{Thomas1994}. By applying a radial confinement, these so
called dust crystals can be used to study many solid state physics phenomena on a scale accessible to ordinary optical
diagnostic techniques.

To understand the behavior of these systems, knowledge about the screening length and the dust charge is of paramount importance. Furthermore, once the charge on the dust particles is known, the
vertical force balance between gravity and the electric field can provide insight into the structure of the electric field in
the sheath. This is important for both dusty plasma experiments and many plasma enhanced industrial processes, where
ions are accelerated by the sheath electric field from the plasma bulk to a processing substrate \cite{Liebermanbook}.

Different techniques using dust particles suspended in the sheath of a plasma as probes to measure properties of the plasma
have been developed in the past. Some techniques depend on measuring the equilibrium levitation height of a few dust
particles, while using passive emission spectroscopy, or probe measurements to obtain additional parameters \cite{Samarian2001, Jay}. Other
techniques use perturbations of the plasma by adding a low-frequency sinusoidal signal to the lower electrode, or sudden
changes in the DC bias \cite{Tomme2000}, or perturb the dust particles by using lasers to excite dust oscillations
\cite{Homann1999}.

Despite being successful, these methods do have several difficulties; often only a few dust particles are used at a time, limiting
the spatial range of measurements at a given time, optical measurements in the sheath are technically challenging in the
small volume close to the highly reflecting electrode \cite{Huebner2009}, while optical access can be hard to obtain
altogether \cite{Mahony1997}. Probe measurements are usually too perturbative to dusty plasma to be reliable, while induced perturbations to 
the plasma inevitably cause changes to the dust parameters, such as the charge.

By manipulating a two-dimensional dust crystal as a whole with thermophoresis by heating and cooling the lower electrode
\cite{Rothermel2002} and by measuring global observable quantities from top-view pictures, it is possible to reconstruct the electric
field in the sheath by using an analytical model \cite{Hebner2002}. Here, we present the experimental
reconstruction of the sheath electric field above the lower powered electrode in a modified GEC cell
\cite{Hargis1994}. One of our conclusions is that the dust is mostly levitated on the plasma side of the classical Bohm point, consistent with previous
findings.

\section{Experimental setup}\label{sec:setup}

A standard 13.56 MHz radio-frequency (RF) discharge in argon was created in a GEC cell, which is modified for dusty plasma experiments. 
Cover plates with cylindrically shaped \emph{cutouts} milled in them can
be attached on top of the powered bottom electrode. These cutouts provide an electric field that confines the
dust particles radially. The upper electrode is a hollow grounded cylinder, allowing optical access to the dust
crystal from above. Above this electrode, dust containers are installed, which release dust particles when tapped. The
particles then fall down through the hollow electrode into the plasma, until they reach their equilibrium levitation height.

A camera mounted above the
hollow upper electrode is used to provide top-view pictures of the horizontal dust crystal structure. A second camera installed in front of a
side-port takes side-view pictures of the vertical structure of the dust crystal. Two red diode-lasers with
wavelengths of
640 nm and 686 nm are equipped with cylindrical lenses
that shape the laser beams into thin laser sheaths. One sheath is oriented vertically, the other horizontally. The light
scattered by the dust particles is captured by either the top-view or side-view camera. Filters can be used with the
cameras to only allow light at the laser wavelength to be observed, rejecting the plasma glow. Typically, only one camera-laser
combination is used at a time. The cameras and lasers can be moved in three directions; however, the focus of
the cameras is
fixed, so that the resolution does not change between pictures. The resolution of the top-camera and side-camera
were measured to be 26 and 14 $\mu$m per pixel, respectively.

\begin{figure}[!ht]
\centering
\includegraphics[width=2.5in]{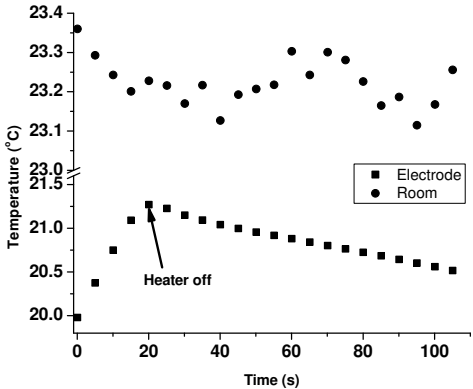}
\caption{The room-temperature and the temperature of the powered electrode measured with thermoresistors. The
room-temperature remains close to 23${}^{\circ}$C. The temperature on the electrode was set at 20${}^{\circ}$ C. It can be seen that the heater/thermostat rapidly heats the system, but then overshoots by a
few degrees. After reaching the maximum temperature, the electrode slowly cools down, giving a relatively long time-window
where the temperature may be considered constant.}
\label{fig:temperaturechange}
\end{figure}

The lower electrode can be cooled or heated by running liquid through a circular channel in the electrode. For heating and cooling, we used a MGW Lauda Brinkman RM3T chiller/heater, connected to a
Ranco ETC-111000-000 thermostat. The temperature was measured with thermoresistors attached near the thermostat, as well as on the
bottom of the electrode. Due to the hysteresis of the thermostat/chiller-heater system, the temperature
fluctuated approximately 2${}^{\circ}$ C above and below the set temperature. A rapid rise to temperatures above
the set temperature was followed by a slow decrease to a point 2 degrees below the set-point. By measuring the electrode-temperature
variation, we determined an interval of at least 60 seconds where the temperature change was slow and the temperature could
be considered constant. This was long enough to take the necessary
top- and side-view pictures, since the cameras operated at 60 and 30 Hz, respectively. Figure \ref{fig:temperaturechange}
shows the change in temperature of the lower electrode and the ambient temperature. 

To determine the electric field
profile, the temperature of the lower electrode was varied from 0${}^{\circ}$ C to 60${}^{\circ}$ C in steps of 10${}^{\circ}$ C. Since
the electrode and the cover-plate are both made of metal and are heated relatively
homogeneously, we expect temperature gradients along the surface of the electrode to be small.
We are therefore not concerned with thermal creep effects recently reported to play an important role in dusty plasma
contained within glass walls \cite{Mitic2008}.

For the experiments presented here, 8.89 $\mu$m diameter melamine-formaldehyde (MF) particles were used and the
discharge was run at 2 W of input power and 200 mTorr background pressure. We used a 1-inch (25.4 mm) diameter cutout, being
roughly 0.5 mm deep.

\section{Observables}\label{sec:Observables}

We assume that the radial confinement potential provided by the cutout is parabolic, so that we can write the potential as 

\begin{equation}
\phi(r,z) = \phi(z-h(r)), ~~h(r) = cr^2,
\end{equation}

\noindent which can be visualized as vertically stacked parabolically shaped equipotential surfaces with the central axis of
the cutout (with $r=0$) as the symmetry-axis. For particles larger than 4 micron in diameter the ion drag can be ignored in
the vertical force balance \cite{Hebner2002}, so that we have 

\begin{equation}
-m_Dg - q_D \left|\frac{d\phi(r,z)}{dz}\right| + F_{th} = 0,
\end{equation}

\noindent where $m_D$ is the dust mass, which we assume to know, $g$ the magnitude of the gravitational acceleration, $q_D$ the dust charge.
The induced thermophoretic force is given by

\begin{equation}
F_{th} = -\frac{32}{15} \frac{{\kappa}_T R^2}{v_T}\nabla T_{gas},
\end{equation}

\noindent where ${\kappa}_T = 0.0177$ is the argon heat conduction coefficient, $R$ the particle radius, $V_T$ the thermal
velocity of the neutral atoms and $\nabla T_{gas}$ the gradient in temperature of the gas between the electrodes. For the
radial confinement we have

\begin{equation}
-q_D \frac{d\phi(r,z)}{dr} = -k r.
\end{equation}

\noindent By using partial differentiation and combining these equations we find the equation for $k$, 

\begin{equation}\label{eq:k}
k = 2c\left[m_Dg-F_{th}\right],
\end{equation}

\noindent which thus depends on the steepness of the potential well and the applied electrode temperature.

Assuming that our crystal is single-layer and consists of $N$ particles with inter-particle spacing $\Delta$ (which depends on
$r$) interacting through an inter-particle potential $V(\Delta)$, it can be shown \cite{Hebner2002} that the pressure in the
crystal layer can be written as

\begin{equation}
P = -\frac{1}{N}\frac{d\left[3NV(\Delta)\right]}{d(\sqrt{3}{\Delta}^2/2)} -
\frac{\sqrt{3}}{\Delta}\frac{dV(\Delta)}{d\Delta}.
\end{equation}

\noindent Using appropriate boundary conditions, the theoretical edge of the crystal $R_{\infty}$ can be shown to be

\begin{eqnarray}
R_{\infty} = \frac{3}{k}\left(3+\frac{{\Delta}_0}{{\lambda}_D}\right)V({\Delta}_0),
\end{eqnarray}

\noindent where ${\Delta}_0$ is the central inter-particle spacing. The actual observable
outer edge of the crystal, $R_M$, is given by $R_M = R_{\infty} - {\Delta}_M\sqrt{3/2}$, with ${\Delta}_M$ the inter-particle spacing at the edge of the crystal. 
The total number of particles in the layer, $N$ can be derived as

\begin{equation}
N = \frac{2\pi\sqrt{3}}{k{\Delta}_0}\left(\frac{1}{{\Delta}_0} + \frac{1}{{\lambda}_D}\right)V({\Delta}_0).
\end{equation}

Combining the above two equations and solving for $V({\Delta}_0)$, we can obtain an equation for the Debye length,
\begin{equation}\label{eq:lambda}
{\lambda}_D = {\Delta}_0\left[\frac{A-S}{3S-A}\right],
\end{equation}

\noindent where we have defined the total crystal surface area as $A = \pi R_{\infty}^2$ and the total surface area covered by $N$
Wigner-Seitz cells measured at the center to be $S = \sqrt{3}{\Delta}_0^2N/2$ After determining the Debye length, the dust charge can be calculated from 

\begin{equation}\label{eq:charge}
q_D = \sqrt{\frac{4\pi{\epsilon}_0{\Delta}_0kR_{\infty}^2}{3\left(3+\frac{3S-A}{A-S}\right)\exp(\frac{A-3S}{A-S})}}.
\end{equation}

Once we have independently determined $k_0$, the constant of the radial confining force without applied thermophoresis, we
can calculate $c$ and then we can calculate $k$ for any additionally applied thermophoresis from equation (\ref{eq:k}). After measuring
the observables  $R_M$, $N$, ${\Delta}_0$, and
${\Delta}_M$, we can determine the Debye length and the dust charge from equation (\ref{eq:lambda}) and equation
(\ref{eq:charge}), respectively. When we obtain the levitation height of the crystal from
a side-view picture, we can then use the force balance to reconstruct the electric field as a function of height.

\subsection{Measuring $k$ without thermophoresis, $k_0$}

The common assumption concerning the radial confinement resulting from cutouts or other adaptations to the lower electrode is that
it can be described by a potential well varying quadratically with the distance from the center. Even though this has been shown
for specific experiments \cite{Konopka2000}, this is not \textit{a priori} clear for our case.

\begin{figure}[!ht]
\centering
\includegraphics[width=2.5in]{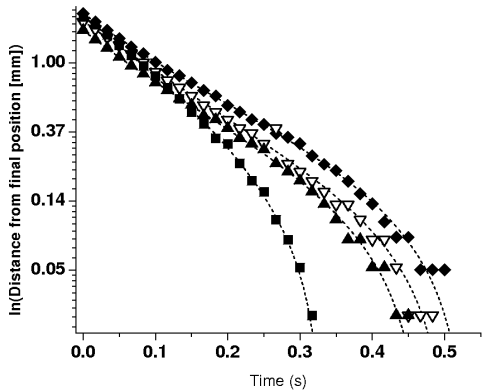}
\caption{The natural logarithm of the distance from the position in the final frame plotted versus time for four selected
particle trajectories. The dashed lines are fits ($R^2 > 0.99$)
according to the solution of equation (\ref{eq:general}).}
\label{damped_motion}
\end{figure}

In order to determine the radial restoring force, we ran the discharge at high pressure (1 Torr in this case) in order to
minimize any oscillatory motion of the dust particles, without heating or cooling of the lower elecotrode, so that $F_{th} = 0$, and dropped the 
dust particles just outside of the cutout. We then monitored the particles while they moved inwards into the potential well of
the cutout. Assuming that in this case the forces acting on the particles are the electrostatic force (determined by the potential well)
and the gas drag, the equation of motion for these particles can be written as:

\begin{equation}\label{eq:overdamped}
m_D\ddot{x} + \alpha\dot{x} + k_0 x = 0,
\end{equation}

\noindent with $\alpha$ the Epstein neutral drag friction coefficient \cite{Epstein}. We assume that the particles are monodisperse, so that they
all have the same mass, $m_D$. 

The general solution to equation (\ref{eq:overdamped}) for the overdamped case (when ${\alpha}^2 > 4 m_D k_0$) is given by a sum of two exponentials,

\begin{equation}\label{eq:general}
x(t) = A \exp({C_1 t}) + B \exp({C_2 t}),
\end{equation} 

\noindent with $C_{1,2}$ given by

\begin{equation}
C_{1,2} = \frac{1}{2} \left[-\frac{\alpha}{m_D} \pm \sqrt{{\left(\frac{\alpha}{m_D}\right)}^2-4\frac{k_0}{m_D}}\right].
\end{equation}

\noindent Taking their sum and difference we find

\begin{eqnarray}
C_1 + C_2 &=& -\frac{\alpha}{m_D}\\
C_1 - C_2 &=&  \sqrt{{\left(\frac{\alpha}{m_D}\right)}^2 - 4\frac{k_0}{m_D}}.
\end{eqnarray}

\noindent Hence, fitting trajectories of particles moving in the potential well provides us with $C_1$
and $C_2$, from which we derive $\alpha/m_D$ and $k_0/m_D$.

Figure \ref{damped_motion} shows four selected particle trajectories. Plotted is the natural logarithm of the distance from
the final position in the last frame. The dashed lines are fits according to equation (\ref{eq:general}). The fits are very
reliable ($R^2>0.99$). From the fits, we determine $\alpha/m_D = 14\pm 1.9$ s${}^{-1}$ and $k_0/m_D = 6.4 \pm 0.87$
s${}^{-2}$. For our 8.9 micron
MF particles, we have $m_D = 5.55\times{10}^{-13}$ kg, so that we have $\alpha = 7.8 \pm 1.1 \times {10}^{-12}$ kg/s, which is
in reasonable agreement with results reported in \cite{Chu2009}. For $k_0$ we thus find $k_0 = 3.6 \pm 0.5\times {10}^{-12}$
kg/s${}^2$.

\section{Determination of the Debye length, charge, and electric field}

Now that we have determined $k_0$ for the 1-inch diameter cutout in our
modified GEC cell, we can take pictures of the dust crystals formed with
different applied temperature gradients. From side-view pictures the levitation
height was determined, and from top-view pictures the number of
particles, crystal radius, and central inter-particle distance were determined using graphics software \cite{ImageJ}.

\begin{figure}[!ht]
\centering
\includegraphics[width=3.0in]{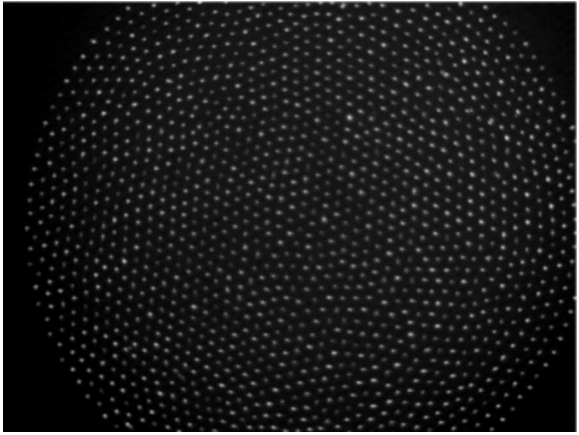}
\caption{A top-view of a dust crystal composed of about 1000 particles 8.9 $\mu$m in diameter, at an electrode temperature of
0${}^{\circ}$ C, applied power of 2 W, and a pressure of 200 mTorr. The
crystal has a radius of 8.2 mm and a central inter-particle distance of 386 $\mu$m. The image has been processed to improve
the visibility of the dust particles.}
\label{crystal_topview}
\end{figure}

Figure \ref{crystal_topview} shows a top-view image of a typical crystal levitated in the discharge. This crystal was formed
while the lower electrode was cooled to 0${}^{\circ}$ C. It contains roughly N=1000
particles, has a radius of $R_{\infty}$= 8.2 mm, and the central
inter-particle distance was measured to be ${\Delta}_0$= 386 $\mu$m. This yields
$A=2.4 \times {10}^{-4}$ m${}^{2}$ and $S=1.3 \times {10}^{-4}$ m${}^{2}$, so that from equation (\ref{eq:lambda}), we
obtain ${\lambda}_D \approx 390 \mu$m.

\begin{figure}[!ht]
\centering
\includegraphics[width=3.0in]{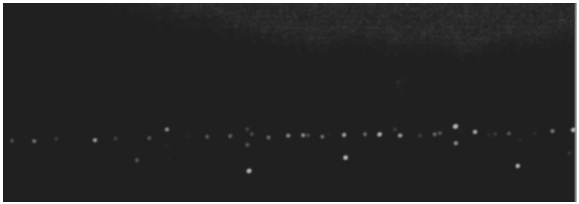}
\caption{A side-view of a dust crystal layer composed of 8.9 $\mu$m diameter particles. It can be seen that the crystal levitated in the
sheath is not completely single-layer. The gray area in the upper-right corner is glow from the plasma. The figure has been
processed to improve the visibility of the dust particles.}
\label{fig:side-view}
\end{figure}

Since the thermophoretic force acts in the same direction as gravity when the electrode is cooled, \textit{k}
increases with decreasing lower electrode temperature. For the applied temperature gradient of 906 K/m, we find that \textit
k (0${}^{\circ}$ C) is 5.3 $\times {10}^{-12}$ kg/s${}^2$. Using this and the value for
${\lambda}_D$ in equation (\ref{eq:charge}), we find a dust charge of $q_D \approx$ -1.3$\times
{10}^{4}$ e. Using the vertical force balance, this results in an electric field of $E = (m_Dg - F_{th})/q_D \approx$ 4600
V/m. From side-view pictures, like the one shown in figure \ref{fig:side-view}, the levitation height of the upper 
crystal-layer was determined to be 5.8 mm. 

\begin{figure}[!ht]
\centering
\includegraphics[width=2.5in]{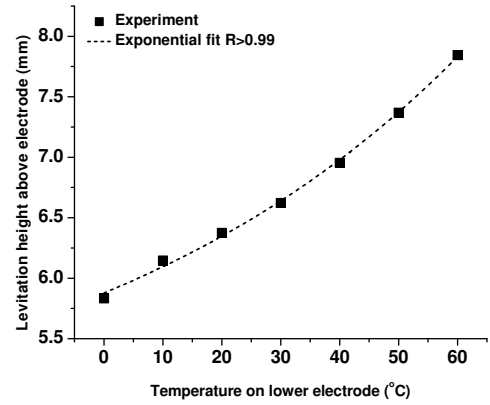}
\caption{The levitation height of the dust crystal versus temperature of the lower electrode, determined from side-view
images. The dashed line is an exponential fit to the data.}\label{fig:lev_height}
\end{figure}

\begin{figure}[!ht]
\centering
\includegraphics[width=2.5in]{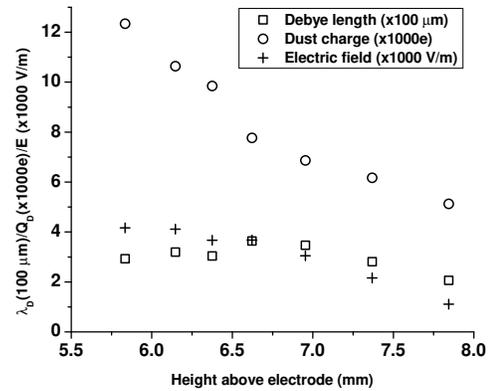}
\caption{The Debye length, dust charge, and electric field plotted versus height above
the electrode, determined from top-view and side-view pictures. The electrode
temperature was varied between 0${}^{\circ}$ C and 60${}^{\circ}$ in steps of 10
degrees.}
\label{exp_results}
\end{figure}

Repeating these calculations for different
applied lower electrode temperatures, the Debye length, dust charge, and electric
field were obtained as a function of the temperature. A plot of the dust levitation height versus applied
temperature is shown in figure \ref{fig:lev_height}. The fit shows an exponential dependence of the levitation height with
the temperature. The Debye length, dust charge, and electric field are shown in figure \ref{exp_results}.

As shown, the Debye length varies between 200 and 360 $\mu$m, having its smallest values higher in the plasma. This seems reasonable,
since the plasma density is expected to increase farther away from the electrode, and ${\lambda}_D \propto \sqrt{T/n}$. Since
the ion temperature is closely coupled to the neutral gas temperature, it has to be mentioned that when the screening is
mainly determined by the ions, that the increase in gas temperature could lead to an increase in ion Debye length. The data
suggest that the decrease in plasma density with height is more important. The measured values for the Debye length are in agreement with values reported in the same GEC cell (roughly 250 $\mu$m for a discharge at 200 mTorr and 5 W
power), measured using a different technique \cite{Jay}, and also with 
the observed inter-particle distances, which in these crystals typically are a few times the Debye
length \cite{Thomas1994}.
 
The dust charge becomes continuously less
negative from approximately -1.3$\times{10}^{4}$e to -5.5$\times{10}^3$e with increasing height above the lower electrode. These values are lower than
the value predicted by OML theory \cite{Allen1992}, which is in agreement with previous results, for instance those reported
in \cite{Trottenberg1995}. 

The electric field decreases approximately linearly for heights 
above 6.5 mm. The obtained vertical electric field in the sheath varies from -4600 V/m (closer to the electrode) to -1380 V/m
(farther up in the plasma). The drift velocity $v_d = {\mu}_{+} E$ for argon ions (with mobility of roughly 1 m${}^2$/Vs
\cite{Frost1957}) then varies from 4600 m/s to 1380 m/s. According to the Bohm criterion, the ions are entering the sheath at the Bohm velocity,
$v_B \equiv \sqrt{k_BT_e/m_{+}}$ m/s, which, for typical electron temperature of 3-5 eV \cite{Hargis1994}, corresponds to
2700-3500 m/s.  We conclude that we are probing the plasma both above the Bohm point, consistent with \cite{Hebner2002}, as
well as just below the Bohm point.

\section{Discussion and conclusions}

There are three main assumptions in the analysis of our measurements. First, we assume that the particles are radially confined in a
parabolic potential well provided by the cutout in the electrode, second, that the ion drag force does not play a role in the
vertical force balance for the 8.9 micron diameter particles levitated in the sheath, and third that the equation of state
derived for a single-layer dust crystal and the resulting equations for the observables are valid in our experiment.

The trajectories of particles moving into the potential well seem to behave very much like an overdamped harmonic oscillator
with a restoring force in Hooke's form, so that $F = - dU(r)/dr = -k r$, hence $U(r) \propto c r^2$. There was no indication of
higher order terms playing a role in the trajectories. This might not be true much closer to the
cutout edge, but plays no role for our current analysis.

Analysis in \cite{Hebner2002} has shown that the ion
drag force needs only to be taken into account for particles smaller than 4 $\mu$m in diameter. Since then, however, the
understanding of the ion drag has improved, for instance where non-linear scattering, ion-neutral collisions, and
anisotropic screening due to ion flow are concerned, which might be important in the sheath \cite{Khrapak2002,Hutchinson2006,
Ivlev2004}. We cannot completely rule out the role of the ion drag, but the values obtained for the charge, Debye length and
the electric field are consistent with values obtained from the literature, so that we expect the role of the ion drag to be
small in our case.

Side-view pictures show that the observed crystals were not completely single-layer, but that layer splitting occurs,
especially at higher electrode temperatures. Defects induced by this process are a source of free energy in the crystal and
therefore the equation of state might not be correct. However, these defects mainly affect the crystal locally. The number of
particles and the crystal radius are global observables, which are probably not affected by these defects. The compression of
the crystal in the parabolic potential well was still observed. When measuring inter-particle distances we took averages over
many different inter-particle distances at various locations. All-in-all, we think that the local deviations from the pure
single-layer crystal state have not affected our results strongly.

The systematic error due to the assumptions in our model are believed to be small, but a word about the precision in
obtaining our observables is then required. Surprisingly, the observable hardest to obtain was the 
total number of particles, both due to the fact that the crystals did not always fit in our field-of-view
while maintaining the resolution necessary to resolve the dust grains, as well as due to the layer-splitting. We
estimate the error to be roughly 10-15\%. The distances measured from the top-view pictures were much more precise. We estimate that any measurement of the radius is off at
most by 3-5 pixels, given our method of fitting a circle to the crystal. This corresponds to roughly 50-80 micron, which for
an 8 mm radius crystal is an error of at most 1\%. The inter-particle distances could be off by a similar number of pixels,
but the results are the averages of many measurements. The error is anticipated to not exceed 5\% because of this. We see
that the largest error in our experiment is actually presented by the determination of $k_0$, which is roughly 14\%.

An interesting observation, which we could not find in the literature, was that a measurable change in the natural DC bias on the
lower electrode occured. This bias became less negative for electrode temperatures above room-temperature, changing from -13 V when the electrode was at
room-temperature to roughly -10.5 V at +60${}^{\circ}$ C. When the electrode was cooled below room-temperature, we did not observe
the opposite effect as clearly. Since a decrease in the negative DC bias causes dust particles to levitate closer to the lower
electrode, negating the effect of the applied temperature, we maintained the DC bias at -13 V for all temperatures using an
external power supply. This insured that the DC bias did not play a role in the levitation of the dust. 

In conclusion, we have succesfully used an entire dust crystal layer as a probe for the electric field in the sheath in front
of the powered electrode in a modified GEC cell, by employing thermophoresis. We obtained the Debye length, the dust charge
and the electric field by simply measuring observables from top-view pictures and by measuring the levitation height from
side-view pictures. This method provides a single, non-perturbative and practical method, which does not require complex
alterations to dusty plasma experiments. We have shown that 8.9 $\mu$m diameter MF particles are mostly suspended on the plasma side
of the Bohm point in our discharge at typical discharge settings.

\section*{Acknowledgment}

This research was made possible by NSF grant PHY-0648869 and NSF CAREER grant PHY-0847127.

\begin{IEEEbiography}[{\includegraphics[width=1in,height=1.25in,clip,keepaspectratio]{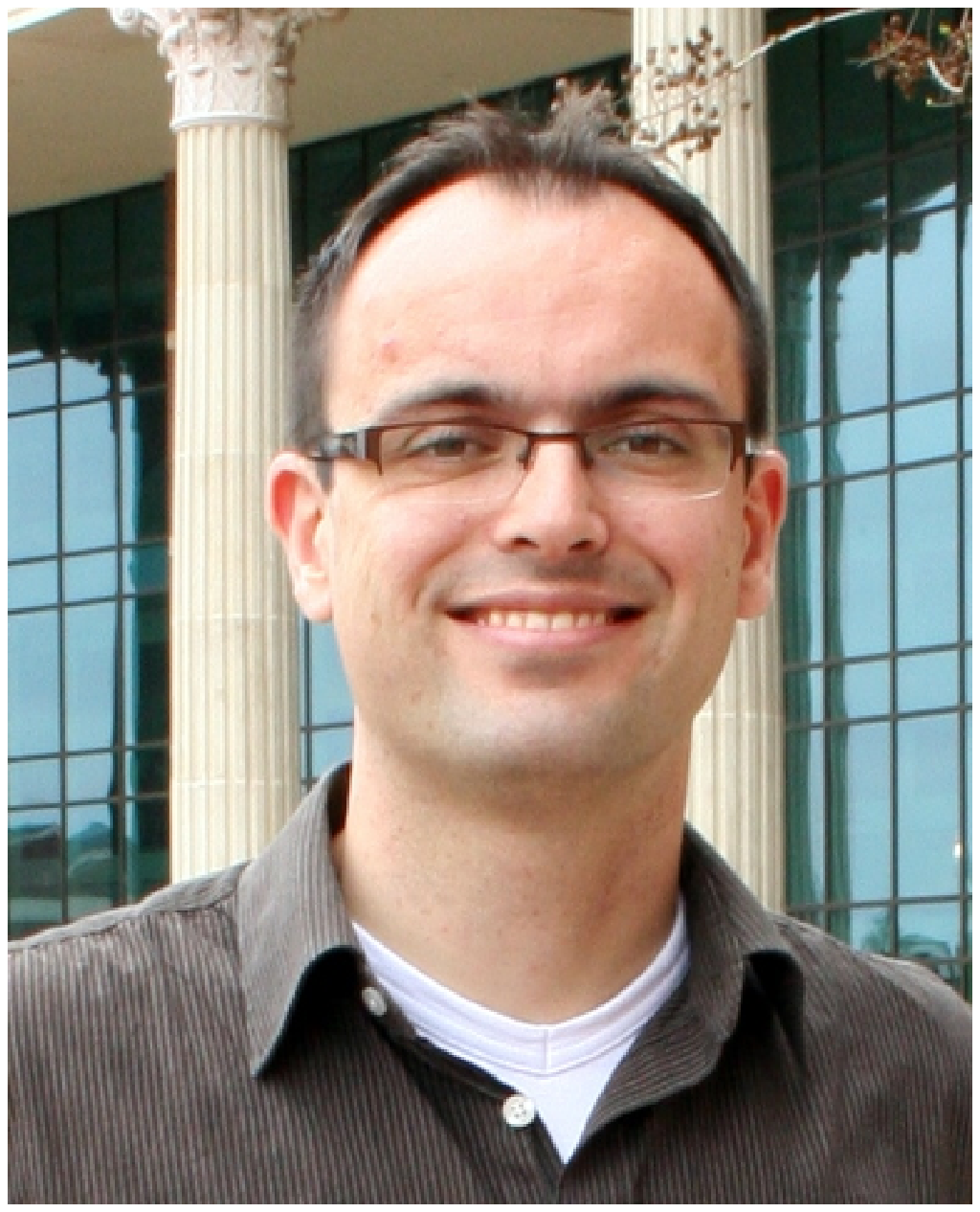}}]{Victor Land} was born in Petten, the Netherlands in 1979 and received his MSc in general astrophysics at Utrecht University in
the Netherlands in 2003, and his PhD at the
FOM-Institute for Plasma Physics 'Rijnhuizen', in the Netherlands in 2007. He is currently a post-doctorate research
associate at the Center for Astrophysics, Space Physics and Engineering Research at Baylor University, in Waco, Texas, where
he works on particle transport, charging, and coagulation in dusty plasma. 
\end{IEEEbiography}

\begin{IEEEbiography}[{\includegraphics[width=1in,height=1.25in,clip,keepaspectratio]{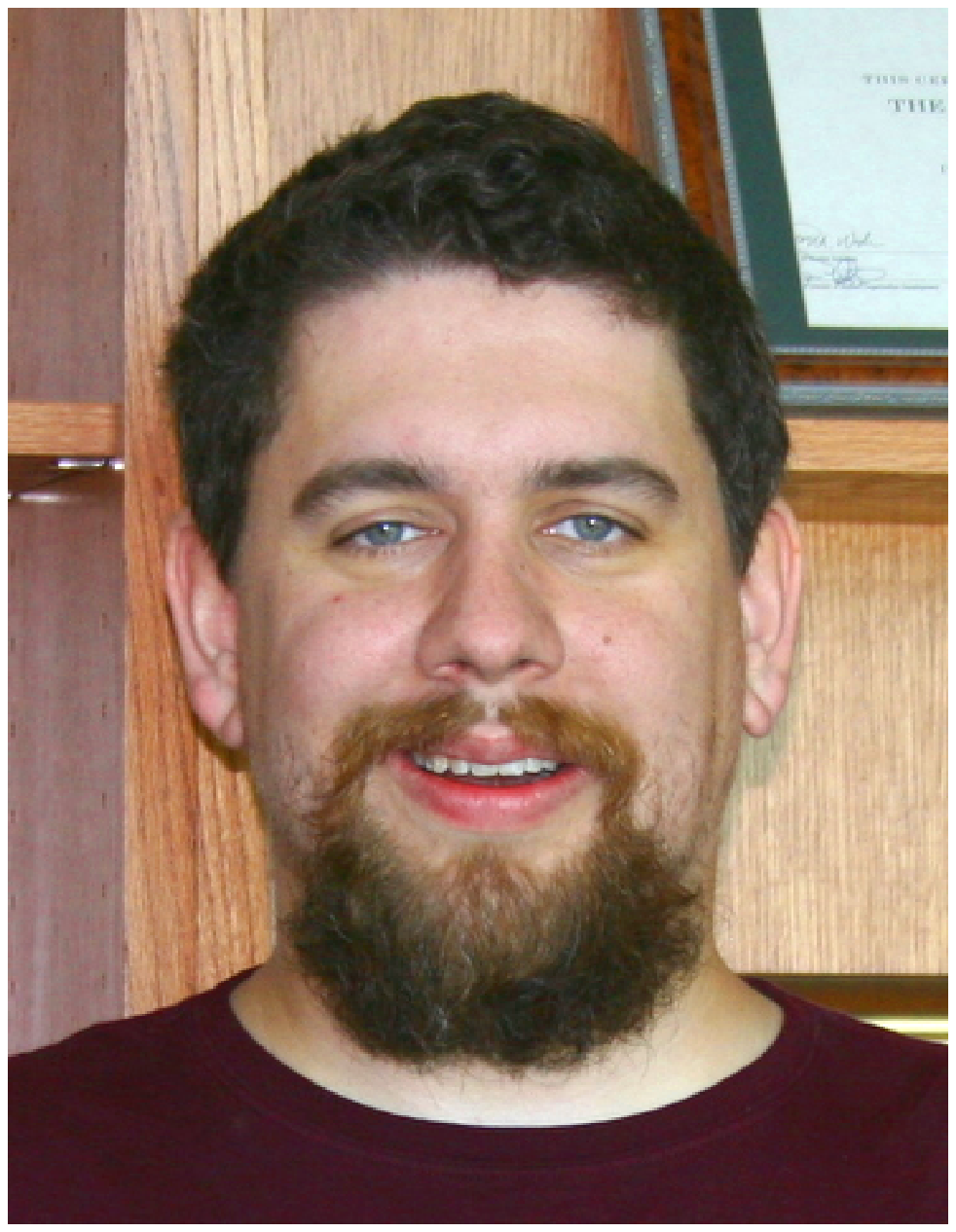}}]{Bernard Smith} was born in Rapid City, 
South Dakota, in 1975.  He received the B.S. degree in physics and math from Baylor University, Waco, Texas, in 1998, the M.S. degree 
in physics from Baylor University in 2003, and the Ph.D. degree in physics from Baylor University in 2005.
His current research interests include complex plasma physics and physics education.
\end{IEEEbiography}

\begin{IEEEbiography}[{\includegraphics[width=1in,height=1.25in,clip,keepaspectratio]{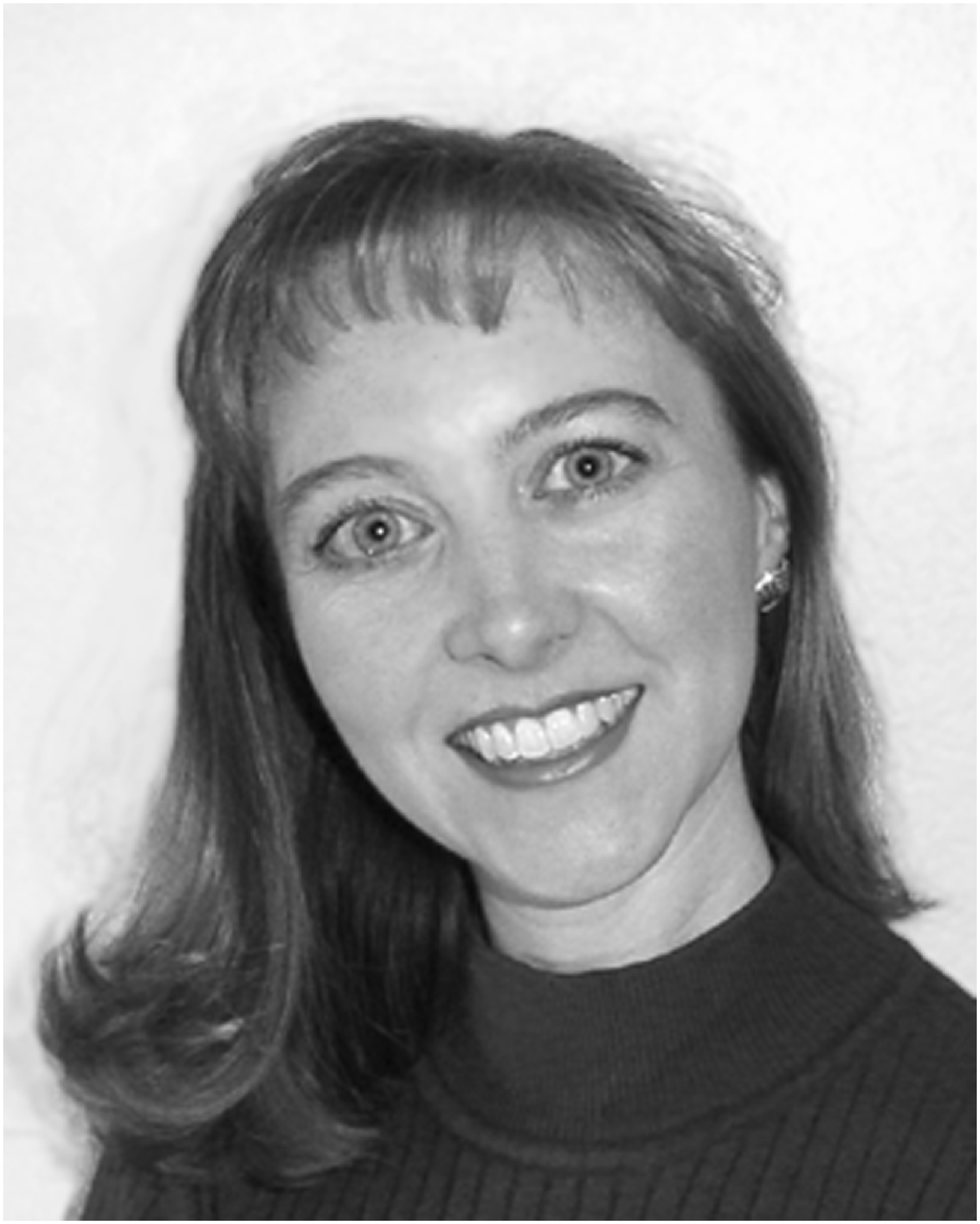}}]{Lorin Matthews}
was born in Paris, TX in 1972. She
received the B.S. and the Ph.D. degrees in physics from
Baylor University in Waco, TX, in 1994 and 1998,
respectively.
She is currently an Assistant Professor in the Physics
Department at Baylor University. Previously, she worked
at Raytheon Aircraft Integration Systems where she was
the Lead Vibroacoustics Engineer on NASA's SOFIA
(Stratospheric Observatory for Infrared Astronomy) project.
\end{IEEEbiography}

\begin{IEEEbiography}[{\includegraphics[width=1in,height=1.25in,clip,keepaspectratio]{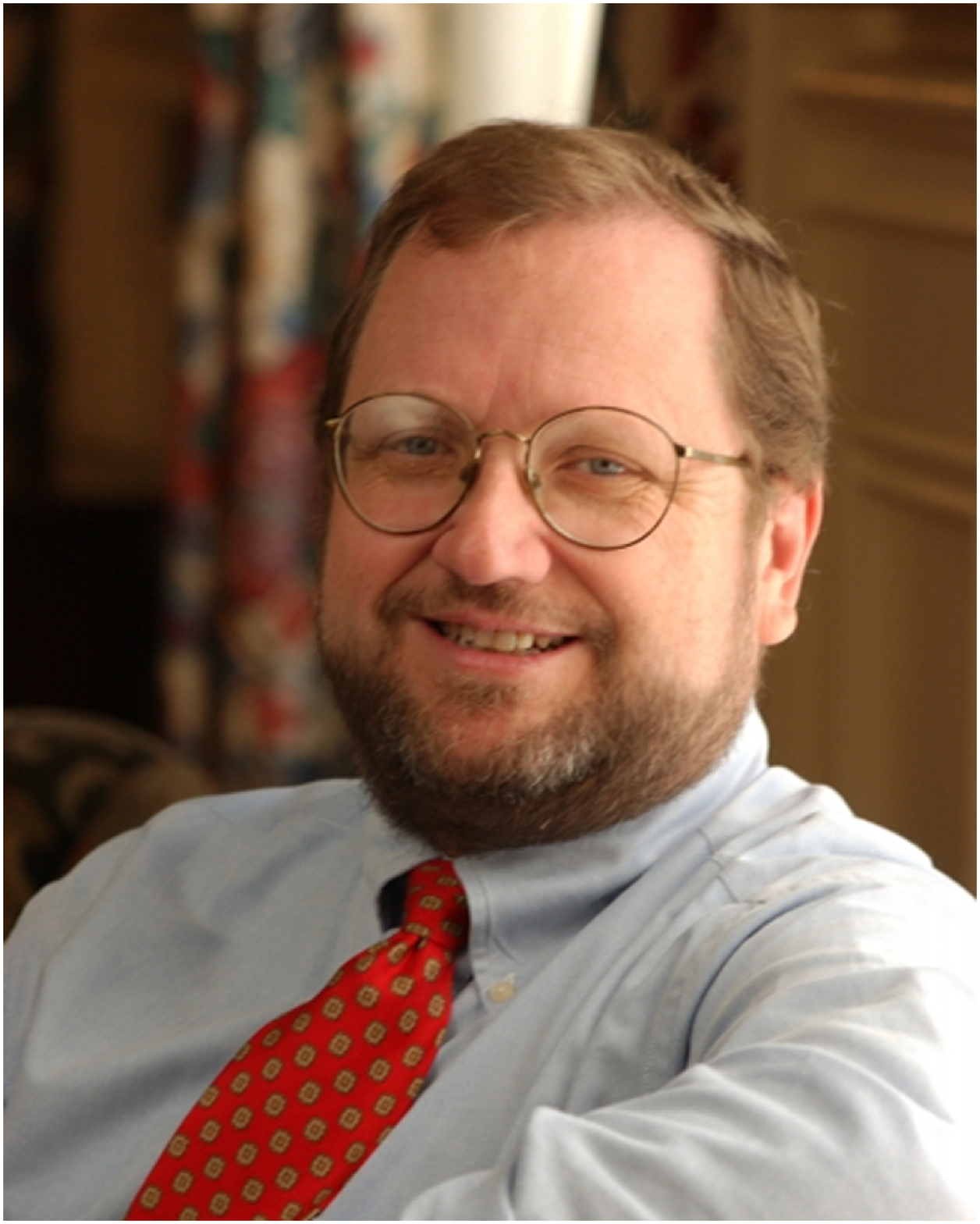}}]{Truell Hyde}
was born in Lubbock, Texas in 1956. He
received the B.S. in physics and mathematics from
Southern Nazarene University in l978 and the Ph.D. in
theoretical physics from Baylor University in 1988.
He is currently at Baylor University where he is the
Director of the Center for Astrophysics, Space Physics \&
Engineering Research (CASPER), a Professor of physics
and the Vice Provost for Research for the University. His
research interests include space physics, shock physics and waves and
nonlinear phenomena in complex (dusty) plasmas.
\end{IEEEbiography}

\end{document}